# Geospatial sensitivity of transmission-constrained ACOPF to generator retirement

Willa Gutowski, Ulyana Shyrokaya, Nicholson Koukpaizan, Joshua Hambrick, Slaven Peles, Eve Tsybina
Oak Ridge National Laboratory, Oak Ridge, TN, USA
Email: {gutowskiwg, shyrokayau, koukpaizannk, hambrickjc, peless, tsybinae} @ ornl.gov

***Abstract* — The US faces a growing resource adequacy challenge: new loads are being added at unprecedented scale while aging generating assets are being retired. In transmission-constrained grids, it is difficult to determine which units can be safely retired and which cannot be retired and instead require lifetime extensions until new generation can be built. Historically, this analysis was prohibitively time consuming. Transmission-constrained AC optimal power flow (ACOPF) is computationally intensive, and a thorough comparison and prioritization of generators could require hundreds or thousands of scenarios. We present an HPC-enabled framework that enables computation and geospatial mapping of the effects of generator retirement in terms of voltage magnitude and angle effects in the steady state. Specifically, our framework detects the effects of generator retirement using a simple k-nearest-neighbors model and a voltage-class-adjusted neighbor model. We demonstrate the results on over 8,000 generator retirement scenarios for a 70,000-bus transmission-constrained synthetic grid.**

## I. Introduction

Rapid growth of the digital economy and the associated datacenter connections have created unprecedented increase in load, placing significant stress on the US electric grid [1]. Simultaneously, generating assets are aging, with many units approaching retirement age. About 360 GW of the country's generating capacity was added during the 1970s and 1980s (Fig. 1), meaning these units will need to be retired or significantly upgraded within the next 5-10 years along the fast load growth. Both the Pennsylvania-Jersey-Maryland Interconnection (PJM) and the Midcontinent Independent System Operator (MISO) recorded a net accredited capacity deficit of approximately 10 GW over 2015-2024 [1], while the North American Electric Reliability Corporation (NERC) long term reliability assessment [2] projects 100 GW in peak seasonal capacity retirement over 2026-2036.

The combination of growing load and retiring equipment creates resource adequacy risks throughout the North American bulk power system. These risks could be mitigated if the retirement of some generating units were deferred until new generation can be built. Moreover, resource adequacy challenges are inherently local: transmission-constrained grids prevent active and reactive power from being transferred to areas with localized reserve shortfalls.

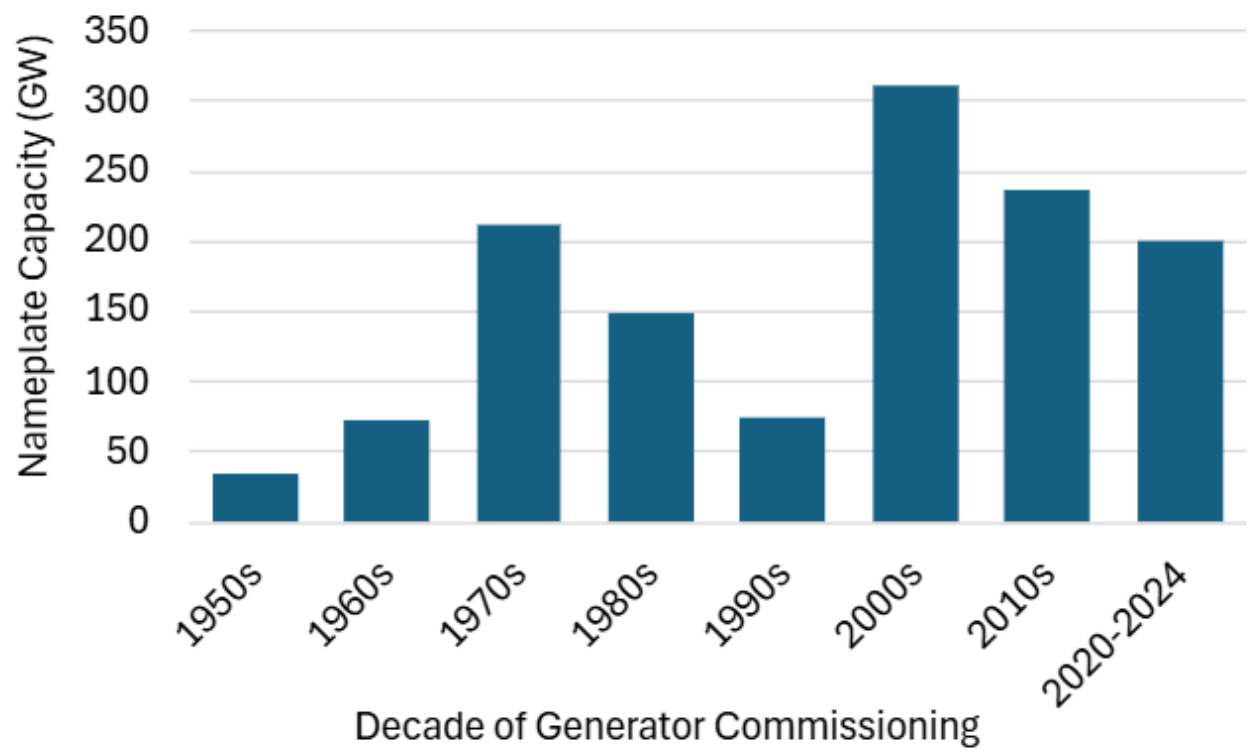

**Fig. 1.** US generating capacity construction by decade [7]

Few studies have examined the effects of generator retirement in transmission-constrained systems [3]-[6]. For example, [3]-[5] identify retirement consequences for system reliability, operating cost, overall fuel mix, and the fuel supply chain. Potential issues that retirement of fossil fuel generators may have in transmission congested grids are discussed in [6]. However, without significant computational resources, these studies are limited to smaller grids and do not address the nationwide planning and coordination challenges facing the US.

Our earlier efforts [8] focused on scanning and prioritizing generators for retirement using fully constrained ACOPF modeling, and was limited to aggregate metrics – overall reliability, system cost, and fuel mix. The retirement of most important units leads to convergence failures as solutions to the ACOPF problem become infeasible, so these are easily flagged as units whose retirement needs to be delayed. The present study extends this work by demonstrating that remaining generators can be evaluated based on the local and regional effects. The retirement of those units will not cause convergence failures but may still cause significant voltage deviations in the nearby buses.

[1]This manuscript has been authored by UT-Battelle, LLC under Contract No. DE-AC05-00OR22725 with the U.S. Department of Energy. The publisher acknowledges the US government license to provide public access under the DOE Public Access Plan (http://energy.gov/downloads/doe-public-access-plan).

We demonstrate the approach on the 70,000-bus synthetic Eastern Interconnection model (ACTIVSg70k) from Texas A&M University (TAMU) [9], with 8,107 candidate generators. For each of them, we demonstrate the effect of the generator retirement on key steady state operating characteristics, voltage magnitude and angle, in 1-10 neighbor bus tiers. The analysis is done for both a simple graph model and through a model which approximates electrical distance between the buses. The results help us understand how far the generator retirement effect propagates through the grid and whether some generators are more impactful than others in maintaining voltage levels far from system limits.

The remainder of this paper is organized as follows: Section II describes the analysis design. Section III presents results and discussion for the models of interest. The summary of our findings and future research are discussed in Section IV.

## II. Analysis and Simulation Design

To address the generator retirement questions, we adopt two modeling approaches. The first is a standard graph model in which all branches are treated equally, regardless of their electrical characteristics, as illustrated in Fig. 2(a). The second approach accounts for voltage-level differences by subdividing neighboring buses into three categories based on voltage class, as illustrated in Fig. 2(b).

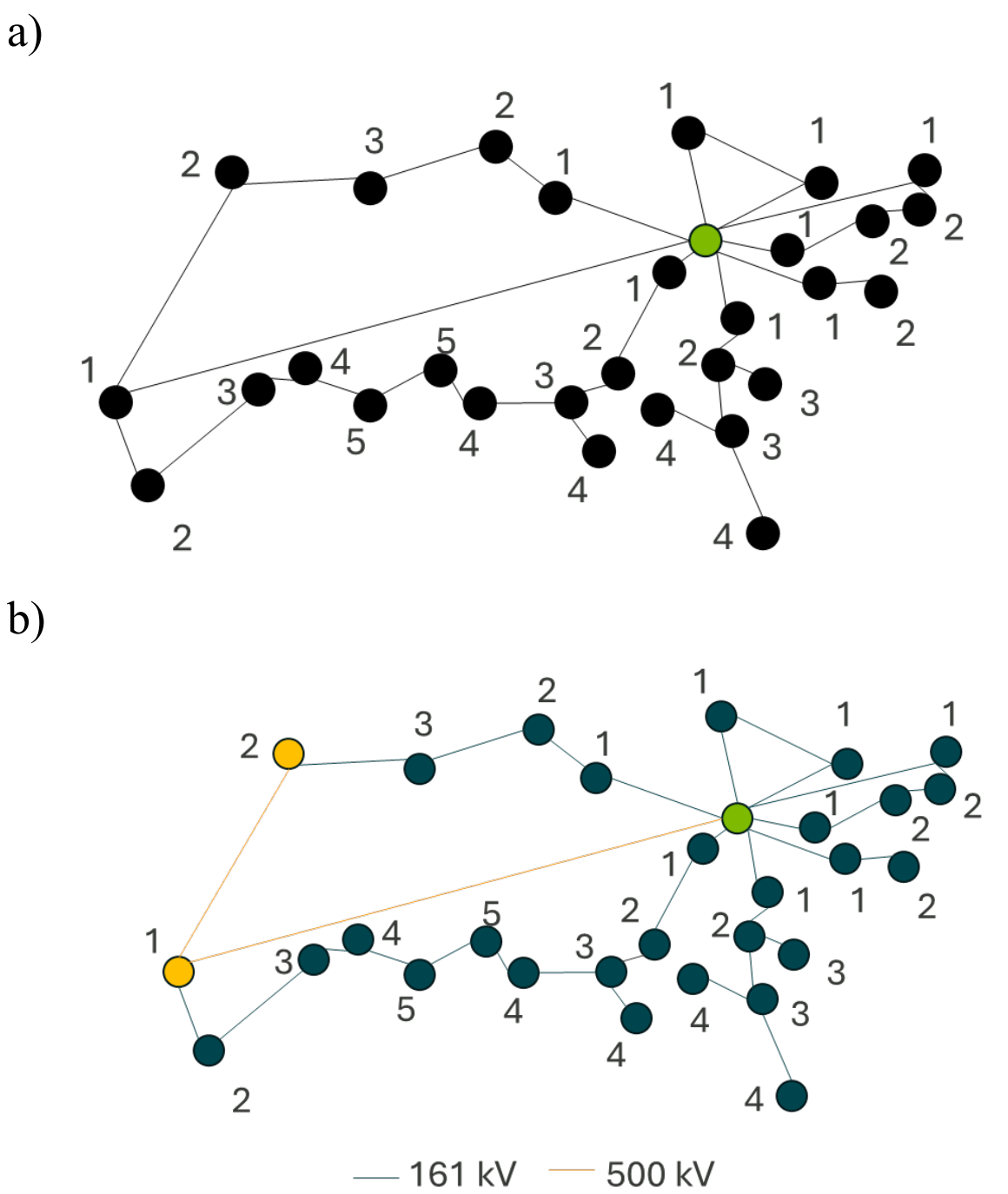

**Fig. 2.** Voltage class logic applied to subdivide neighbors

The first category includes lowest voltage class buses, usually found in sub-transmission and distribution grids (≤ 34 kV). The second category includes the lower end of transmission voltage classes, usually found in regional grids (69 kV to 230 kV). The third category includes buses from voltage classes typical for transmission grids connecting different system areas, such as 345 kV, 500 kV, and 765 kV. Such voltage classes are typically found in large substations with few immediate load connections and potentially with some generation connections. Higher-voltage neighbor buses tend to be geographically more remote and are generally less sensitive to the output of any particular generator.

Once the respective subpopulations of neighbor buses have been identified, we independently disconnect each of the 8,107 operating generators in the system and record nodal voltages across the entire system. We remove the cases that did not converge from the analysis since voltages for those cases will not be representative. For the cases that converge, we calculate voltage deviation as

$$\Delta V_{m,bus} = \frac{V_{m,\text{new}} - V_{m,\text{old}}}{V_{m,\text{old}}}, \% \tag{1}$$

$$\Delta V_{a,bus} = V_{a,\text{new}} - V_{a,\text{old}} \tag{2}$$

Where $V_{\text{m}}$ is voltage magnitude and $V_a$ is voltage angle in each bus of a nodal system. A generator is considered high impact if its retirement causes significant changes in voltage magnitudes or angles as far as three neighbors (PJM standard). We are interested in both magnitude and sign in $\Delta V_m$ and $V_a$. Any generator on the grid can regulate voltage both up and down, causing the generator retirement to cause positive or negative deviations.

The data generation approach follows the methodology described in [8]. We use a modified version of the 70,000-bus TAMU synthetic grid with reduced MVA line limits to induce congestion. Other system characteristics, such as cost curves and generator mechanical constraints, are retained from the original dataset (Fig. 3). The generator profile suggests that a relatively small subset of high-impact units, approximately 500 to 1,000 generators, is capable of causing significant grid disturbances. About half of the units are smaller generators that are unlikely to produce measurable system-wide effects individually; however, even small generators can have outsized impacts in transmission-constrained systems.

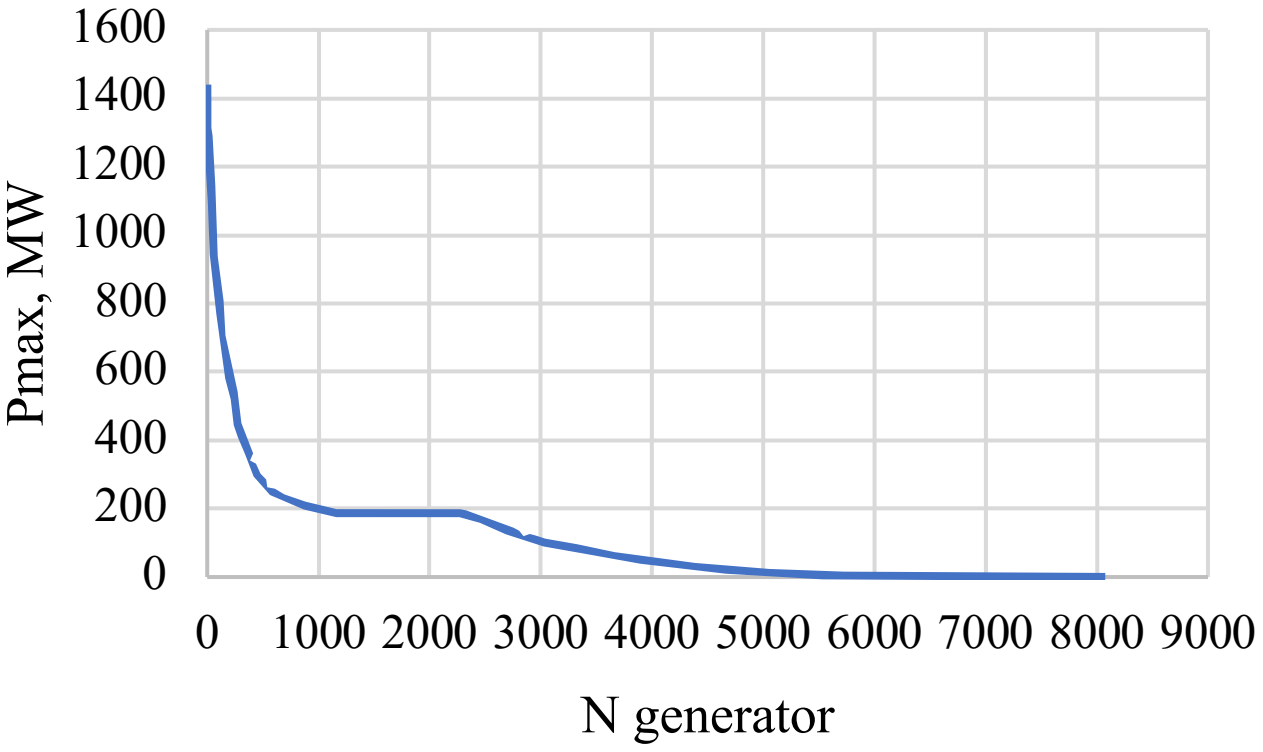

**Fig. 3.** Distribution of size for committed generating units in 70,000-bus TAMU synthetic grid

The three ACOPF scenario sets were processed in parallel batches, each utilizing 427 CPU nodes with 19 scenarios per node. The scenario sets are 1) Main, which enforces MVA limits and does not allow load shedding, 2) Shed Load, which enforces MVA limits and allows load shedding, 3) Unlimited, which does not enforce MVA limit and does not allow to shed load. On average, each scenario required 79.90 seconds to solve using the IPOPT optimization package [10] with the MA57 linear solver [11]. Bus, branch, and generator data were extracted from the individual simulation output files and consolidated into Parquet files ranging in size from 5.8 GB to 77.78 GB. The power flow simulations took 30-214 seconds per scenario, using CPU parallel computations. Impact coefficient calculations, including the SQL aggregation step, completed in about 30 seconds using DuckDB parallelization.

## III. Results and Discussion

### A. Geospatial generator importance for k neighbors

The descriptive statistics for the solutions are provided in Table I. The summary for network connectivity of neighbors by tier is provided in Fig. 4.

Table I. Descriptive statistics

| id_file | Main | Shed Load | Unlimited |
|---|---|---|---|
| Convergence % | 86.58 | 99.89 | 100.00 |
| N cases converged | 7020 | 8099 | 8108 |
| Pg min [GW] | 610.8 | 608.4 | 610.2 |
| Pg mean [GW] | 610.8 | 610.8 | 610.3 |
| Pg max [GW] | 611.5 | 611.3 | 610.6 |
| Qg min [GVar] | 105.4 | 103.2 | 90.6 |
| Qg mean [GVar] | 106.0 | 104.1 | 92.2 |
| Qg max [GVar] | 123.7 | 124.7 | 94.1 |
| System cost min [$10M] | 16.47 | 16.47 | 16.44 |
| System cost mean [$10M] | 16.48 | 16.49 | 16.44 |
| System cost max [$10M] | 16.64 | 18.96 | 16.52 |
| solve_t_min [s] | 68.22 | 79.29 | 31.50 |
| solve_t_mean [s] | 79.90 | 86.19 | 33.72 |
| solve_t_max [s] | 173.46 | 213.80 | 39.92 |

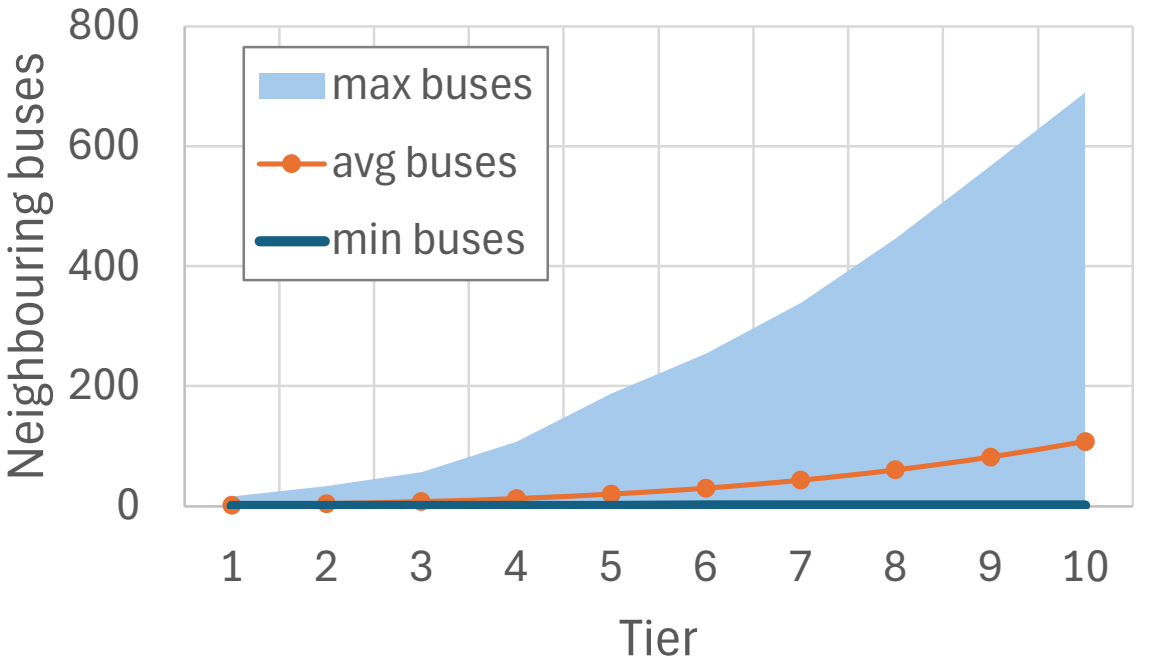

**Fig. 4.** Distribution of the number of neighbors by neighbor tier in the 70,000-bus TAMU synthetic grid.

The average impact of retiring a generator on voltage magnitude was found to be between 0.129% and 0.469% in tier 1 (Fig. 5a). This number predictably declines as the buses get further away from the generator. Tier 1 buses typically represent transmission substations that route power from the generator to other substations in the broader network. Tiers 2 and 3 generally correspond to other buses within the regional network, which depend heavily on the retired generator for both active power delivery and reactive power support through voltage setpoints. More remote neighbors at tiers 4 and 5 exhibit weaker dependency on the retired generator, with a relatively constant impact magnitude, suggesting that these buses belong either to the same regional network or to distant nodes in the higher-voltage transmission backbone. A more detailed breakdown by voltage class is provided in the following subsection. A clear attenuation in impact begins at tier 6, where buses are sufficiently remote, both geographically and electrically, and are likely influenced by other generators. No substantial change in behavior was observed across tiers 6 through 10, indicating that the generator’s effect at these distances dissipates into background variation.

Voltage angle changes, measured in degrees, also decreased as neighbor tier increased, but the scenario ordering differed from the voltage magnitude results. The Unlimited case produced the largest mean absolute voltage angle deviations, whereas the Main case produced the smallest deviations (Fig. 5b).

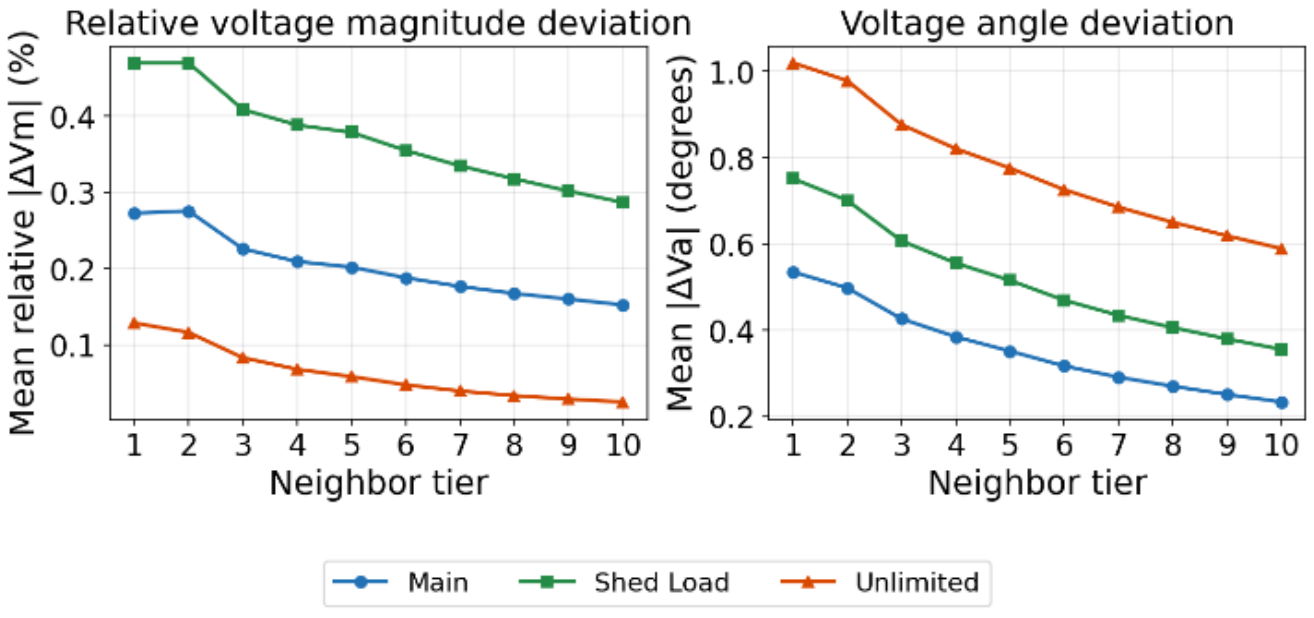

**Fig. 5.** Mean absolute voltage (a) magnitude and (b) angle as a function of simple adjacency

Most generators in the 70,000-bus TAMU grid regulated voltage up, which caused a subsequent drop in voltages after a generator was retired (Fig. 6). Only approximately 30% of generator retirements produced a higher voltage profile on neighboring buses. While voltage regulation can be partially provided by other support devices, this finding confirms that generator retirement may reduce the availability of ancillary services on the grid.

Generators were found to vary substantially in their impact on the grid (Fig. 6). The majority produced a voltage change of at most 2%, indicating that the grid was not significantly weakened by the retirement of those units. However, retirement of some generators caused voltage deviations of up to 13%. These high-impact units appear in Main and Shed Load scenarios, but their number and the magnitude of the associated voltage changes both decline in the Unlimited scenario. Retiring such generators poses a risk to grid steady-state operation. They need to be tested more heavily for contingencies and may be unsafe to retire.

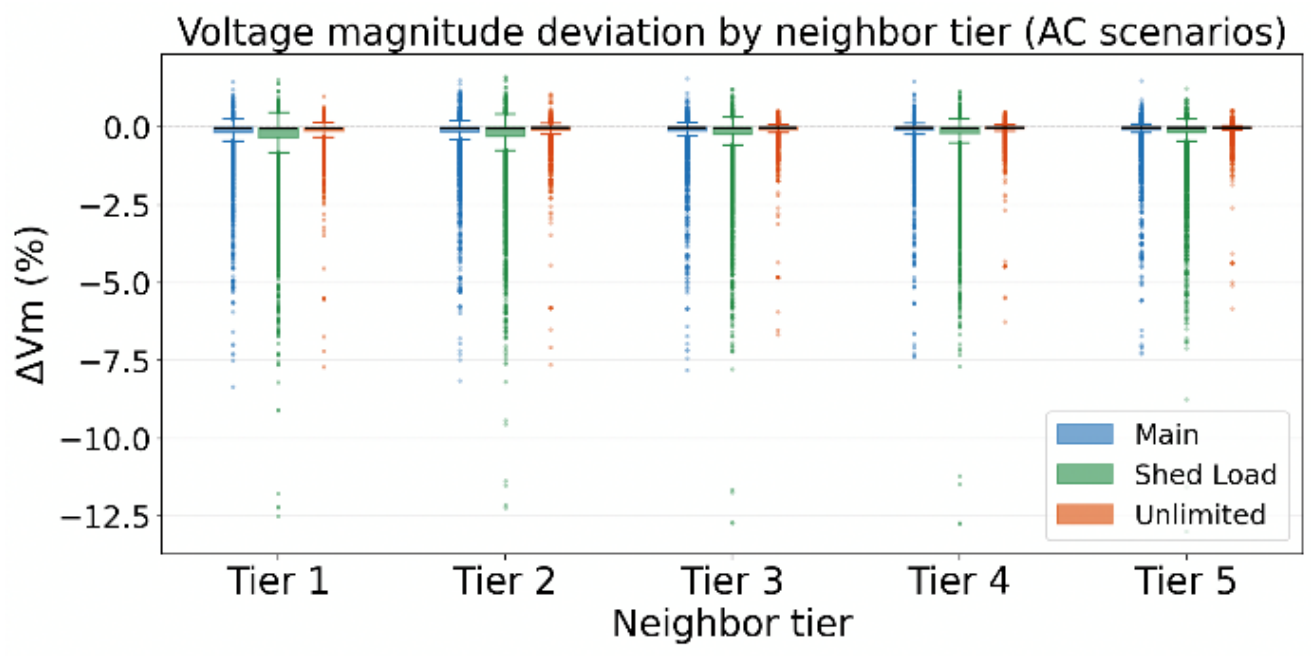

**Fig. 6.** Box plot of generator contingency impact on voltage magnitude for tiers 1-5

Using our methodology, we are able to rank generators that are most impactful if retired, across a range of neighbor bus tiers. Specifically, in the Main case the most impactful generators were generator 1 in bus 29674, generator 1 in bus 28483, and generator 1 in bus 63343 (Table II).

Table II. Generating units ranked by change in voltage magnitude

| Rank | Bus | Gen | \|ΔVm\| [%] | Pmax | fuel |
|---|---|---|---|---|---|
| 1 | 29674 | 1 | 7.82% | 53.9 | hydro |
| 2 | 28483 | 1 | 7.43% | 33.7 | oil |
| 3 | 63343 | 1 | 7.18% | 80 | IBR |
| 4 | 65810 | 1 | 7.15% | 15 | gas |
| 5 | 21581 | 1 | 6.97% | 5 | IBR |
| 6 | 48477 | 1 | 6.71% | 20 | IBR |
| 7 | 61987 | 1 | 6.39% | 204 | IBR |
| 8 | 32390 | 1 | 5.84% | 39.9 | hydro |
| 9 | 32390 | 2 | 5.84% | 39.9 | hydro |
| 10 | 32390 | 3 | 5.84% | 39.9 | hydro |

The geospatial distribution of impactful generators in the TAMU 70,000 bus system is shown in Fig. 7. Some of the more impactful generators can be found in Texas, Alabama, Florida, and South Carolina.

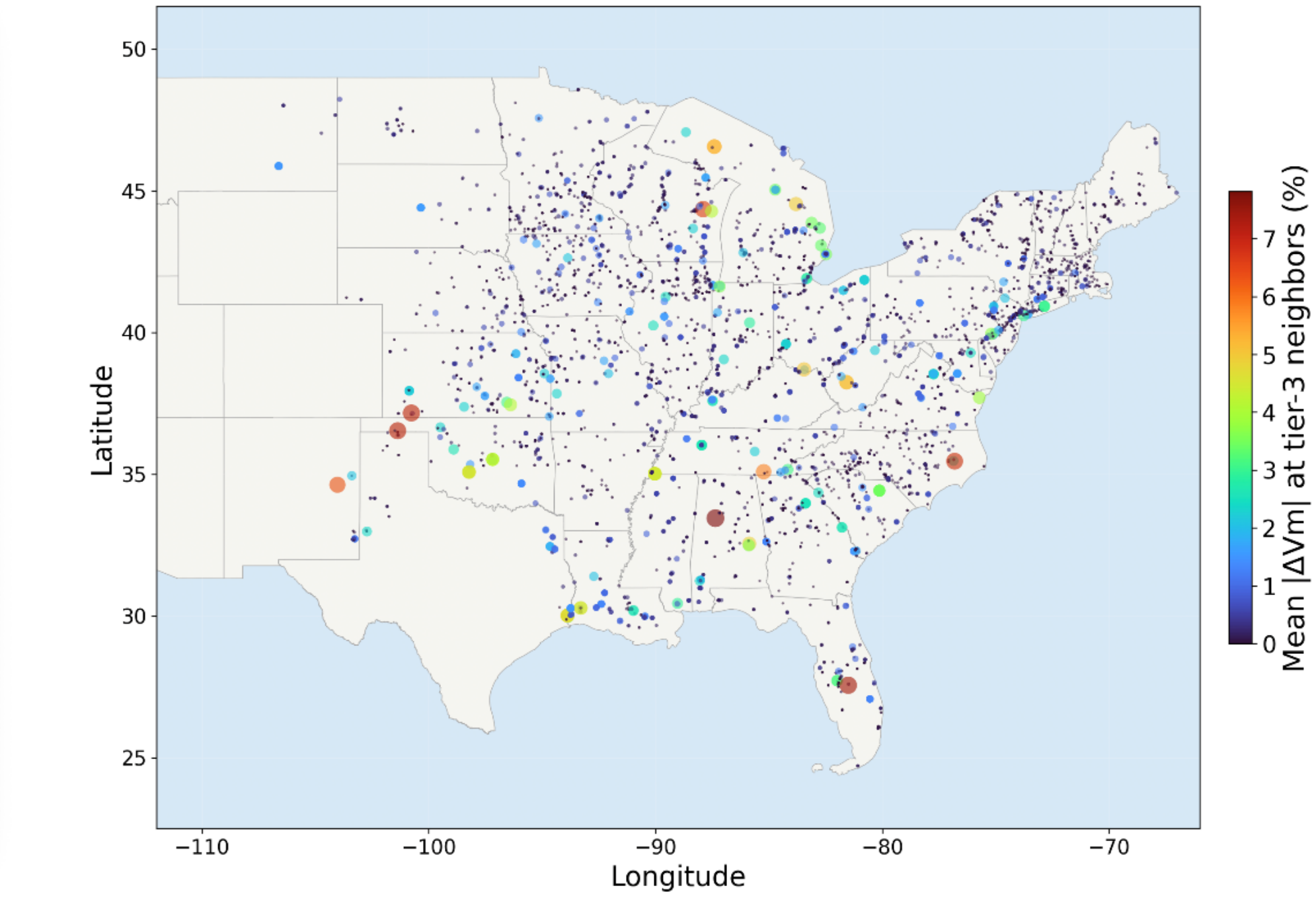

**Fig. 7.** Point map of relative impact of generator retirement on voltage magnitude for synthetic 70,000-bus Eastern Interconnection system.

*B. Generator importance for voltage class adjusted neighbors*

As discussed in the previous subsection, the impact of generators on the grid is affected by voltage and geographical location of the bus. In the following discussion, we further consider the effect of the electrical distance: higher voltage lines tend to be longer, and a bus on a higher voltage grid experiences the tradeoff between higher conductivity and larger distance. In general, buses on higher voltage lines are affected simultaneously by more generators than buses on low voltage lines. Fig. 8 shows the change of generator importance per tier for different voltage classes. The figure shows that, while generators still tend to cluster tightly around ±1-2%, the high impact outliers discussed above are mostly present in lower voltage class grids.

Specifically, distribution voltage buses (below 34 kV) are less prone to generator impact at close tiers because there are fewer such buses on the grid and they are usually located far from generators. Predictably, the effect of generator retirement increases for those buses at sufficiently large neighbor distance. The effect on lower end of transmission voltage (69-230 kV) remains highest across all three voltage groups, confirming the argument that retirement of generators is to a significant extent a local problem. Generators provide significant volumes of active and, importantly, reactive power to lower voltage buses around them. This happens because generator setpoints tend to target nearby buses and because of higher transmission losses for active power in the lower voltage grid. Finally, the higher voltage group of buses (345-765 kV) tends to have smaller and more stable distribution of voltage deviations across neighbor tiers.

## IV. CONCLUSIONS AND FUTURE WORK

This paper presents an HPC-based framework for rapidly assessing the geospatial importance of individual generating units at system scale.

The methodology evaluates and ranks 8,107 generators in the 70,000-bus TAMU synthetic grid in under 30 minutes. The process includes less than six minutes per run to solve the 24,321 generator retirement cases and 3 baseline scenarios where no generator was retired. Data postprocessing and ranking took about 30 seconds. We demonstrate that comprehensive generation portfolio assessments, historically time-consuming and infrequently performed, can be conducted fast and with fidelity comparable to the existing industry-grade interconnection studies.

As future work, we plan to develop a more systematic way to analyze cases that did not converge. While nonconvergence typically implies solution infeasibility, additional attention is needed to ensure the failure is not caused by numerical artifacts. Furthermore, additional analysis comparing results for the fully constrained, load shedding, and transmission unconstrained solutions could provide valuable information on what power grid reinforcements are needed if a large generator is to be retired. We also plan to evaluate multiple unit retirement scenarios, modeled as N-k contingencies.

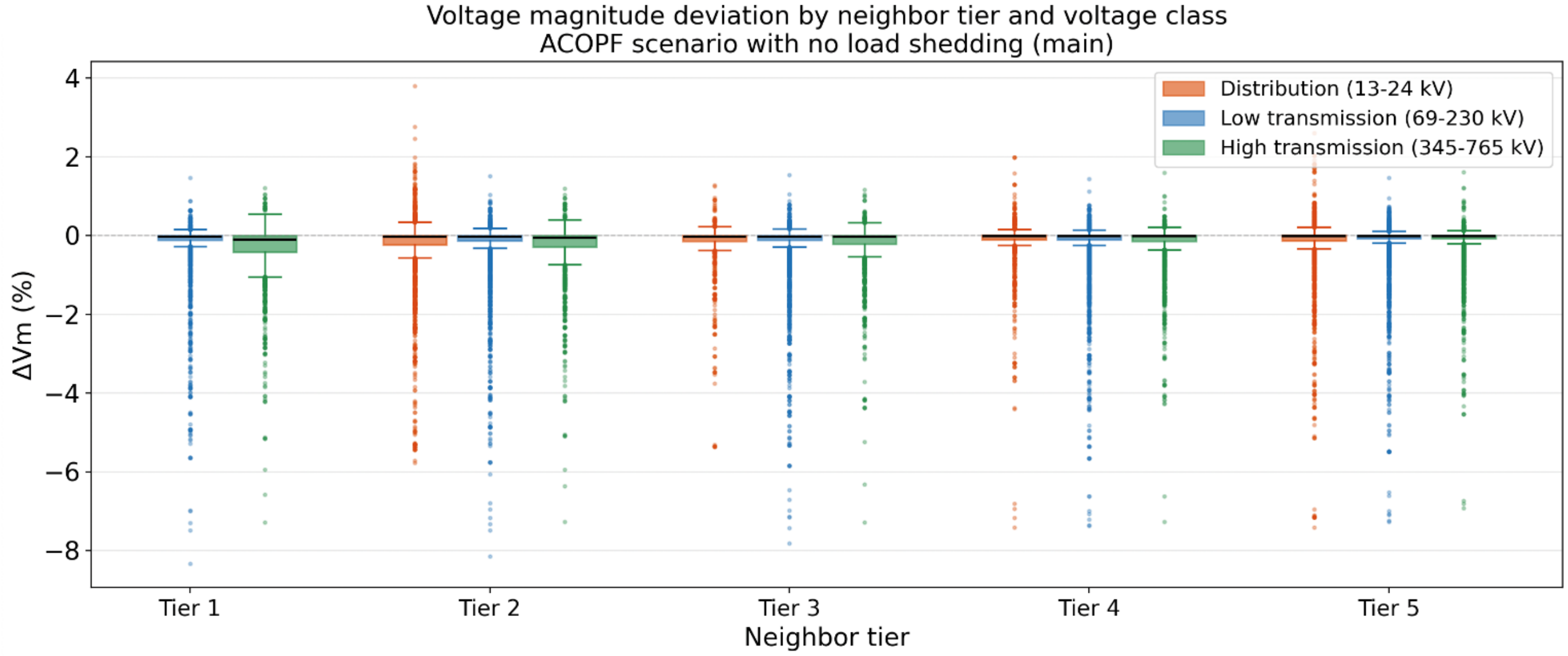

**Fig. 8.** Box plot of generator retirement impact across neighbor tiers 1-5 and voltage class

ACKNOWLEDGEMENTS

This research used resources of the Oak Ridge Leadership Computing Facility at the Oak Ridge National Laboratory, which is supported by the Advanced Scientific Computing Research programs in the Office of Science of the U.S. Department of Energy under Contract No. DE-AC05-00OR22725.